\documentclass[]{eptcs}
\providecommand{\event}{WoC 2015}

\usepackage{tikz}
\usepackage{theorem}
\usepackage{amsmath}
\usepackage{amssymb}

\newdimen\proofrulebreadth \proofrulebreadth=.05em
\newdimen\proofdotseparation \proofdotseparation=1.25ex
\newdimen\proofrulebaseline \proofrulebaseline=2ex
\newcount\proofdotnumber \proofdotnumber=3
\let\then\relax
\def\hfi{\hskip0pt plus.0001fil}
\mathchardef\squigto="3A3B
\newif\ifinsideprooftree\insideprooftreefalse
\newif\ifonleftofproofrule\onleftofproofrulefalse
\newif\ifproofdots\proofdotsfalse
\newif\ifdoubleproof\doubleprooffalse
\let\wereinproofbit\relax
\newdimen\shortenproofleft
\newdimen\shortenproofright
\newdimen\proofbelowshift
\newbox\proofabove
\newbox\proofbelow
\newbox\proofrulename
\def\shiftproofbelow{\let\next\relax\afterassignment\setshiftproofbelow\dimen0 }
\def\shiftproofbelowneg{\def\next{\multiply\dimen0 by-1 }%
\afterassignment\setshiftproofbelow\dimen0 }
\def\setshiftproofbelow{\next\proofbelowshift=\dimen0 }
\def\setproofrulebreadth{\proofrulebreadth}

\def\prooftree{
\ifnum  \lastpenalty=1
\then   \unpenalty
\else   \onleftofproofrulefalse
\fi
\ifonleftofproofrule
\else   \ifinsideprooftree
        \then   \hskip.5em plus1fil
        \fi
\fi
\bgroup
\setbox\proofbelow=\hbox{}\setbox\proofrulename=\hbox{}%
\let\justifies\proofover\let\leadsto\proofoverdots\let\Justifies\proofoverdbl
\let\using\proofusing\let\[\prooftree
\ifinsideprooftree\let\]\endprooftree\fi
\proofdotsfalse\doubleprooffalse
\let\thickness\setproofrulebreadth
\let\shiftright\shiftproofbelow \let\shift\shiftproofbelow
\let\shiftleft\shiftproofbelowneg
\let\ifwasinsideprooftree\ifinsideprooftree
\insideprooftreetrue
\setbox\proofabove=\hbox\bgroup$\displaystyle 
\let\wereinproofbit\prooftree
\shortenproofleft=0pt \shortenproofright=0pt \proofbelowshift=0pt
\onleftofproofruletrue\penalty1
}

\def\eproofbit{
\ifx    \wereinproofbit\prooftree
\then   \ifcase \lastpenalty
        \then   \shortenproofright=0pt  
        \or     \unpenalty\hfil         
        \or     \unpenalty\unskip       
        \else   \shortenproofright=0pt  
        \fi
\fi
\global\dimen0=\shortenproofleft
\global\dimen1=\shortenproofright
\global\dimen2=\proofrulebreadth
\global\dimen3=\proofbelowshift
\global\dimen4=\proofdotseparation
\global\count255=\proofdotnumber
$\egroup  
\shortenproofleft=\dimen0
\shortenproofright=\dimen1
\proofrulebreadth=\dimen2
\proofbelowshift=\dimen3
\proofdotseparation=\dimen4
\proofdotnumber=\count255
}

\def\proofover{
\eproofbit 
\setbox\proofbelow=\hbox\bgroup 
\let\wereinproofbit\proofover
$\displaystyle
}%
\def\proofoverdbl{
\eproofbit 
\doubleprooftrue
\setbox\proofbelow=\hbox\bgroup 
\let\wereinproofbit\proofoverdbl
$\displaystyle
}%
\def\proofoverdots{
\eproofbit 
\proofdotstrue
\setbox\proofbelow=\hbox\bgroup 
\let\wereinproofbit\proofoverdots
$\displaystyle
}%
\def\proofusing{
\eproofbit 
\setbox\proofrulename=\hbox\bgroup 
\let\wereinproofbit\proofusing
\kern0.3em$
}

\def\endprooftree{
\eproofbit 
  \dimen5 =0pt
\dimen0=\wd\proofabove \advance\dimen0-\shortenproofleft
\advance\dimen0-\shortenproofright
\dimen1=.5\dimen0 \advance\dimen1-.5\wd\proofbelow
\dimen4=\dimen1
\advance\dimen1\proofbelowshift \advance\dimen4-\proofbelowshift
\ifdim  \dimen1<0pt
\then   \advance\shortenproofleft\dimen1
        \advance\dimen0-\dimen1
        \dimen1=0pt
        \ifdim  \shortenproofleft<0pt
        \then   \setbox\proofabove=\hbox{%
                        \kern-\shortenproofleft\unhbox\proofabove}%
                \shortenproofleft=0pt
        \fi
\fi
\ifdim  \dimen4<0pt
\then   \advance\shortenproofright\dimen4
        \advance\dimen0-\dimen4
        \dimen4=0pt
\fi
\ifdim  \shortenproofright<\wd\proofrulename
\then   \shortenproofright=\wd\proofrulename
\fi
\dimen2=\shortenproofleft \advance\dimen2 by\dimen1
\dimen3=\shortenproofright\advance\dimen3 by\dimen4
\ifproofdots
\then
        \dimen6=\shortenproofleft \advance\dimen6 .5\dimen0
        \setbox1=\vbox to\proofdotseparation{\vss\hbox{$\cdot$}\vss}%
        \setbox0=\hbox{%
                \advance\dimen6-.5\wd1
                \kern\dimen6
                $\vcenter to\proofdotnumber\proofdotseparation
                        {\leaders\box1\vfill}$%
                \unhbox\proofrulename}%
\else   \dimen6=\fontdimen22\the\textfont2 
        \dimen7=\dimen6
        \advance\dimen6by.5\proofrulebreadth
        \advance\dimen7by-.5\proofrulebreadth
        \setbox0=\hbox{%
                \kern\shortenproofleft
                \ifdoubleproof
                \then   \hbox to\dimen0{%
                        $\mathsurround0pt\mathord=\mkern-6mu%
                        \cleaders\hbox{$\mkern-2mu=\mkern-2mu$}\hfill
                        \mkern-6mu\mathord=$}%
                \else   \vrule height\dimen6 depth-\dimen7 width\dimen0
                \fi
                \unhbox\proofrulename}%
        \ht0=\dimen6 \dp0=-\dimen7
\fi
\let\doll\relax
\ifwasinsideprooftree
\then   \let\VBOX\vbox
\else   \ifmmode\else$\let\doll=$\fi
        \let\VBOX\vcenter
\fi
\VBOX   {\baselineskip\proofrulebaseline \lineskip.2ex
        \expandafter\lineskiplimit\ifproofdots0ex\else-0.6ex\fi
        \hbox   spread\dimen5   {\hfi\unhbox\proofabove\hfi}%
        \hbox{\box0}%
        \hbox   {\kern\dimen2 \box\proofbelow}}\doll%
\global\dimen2=\dimen2
\global\dimen3=\dimen3
\egroup 
\ifonleftofproofrule
\then   \shortenproofleft=\dimen2
\fi
\shortenproofright=\dimen3
\onleftofproofrulefalse
\ifinsideprooftree
\then   \hskip.5em plus 1fil \penalty2
\fi
}

\newcommand\senttype{\textbf{Sen}}

\newcommand\newgramsym\overline 
\newcommand\gramrule{\mathrel{{:\mkern-0.8mu:=}}}
\newcommand\subtype{\mathrel{\le}}

\newcommand\hole{\bigcirc}

\newcommand\wand{\mathbin{{-}\mkern-10mu{-}\mkern-6mu{*}}}

\newcommand\lamprod{\mathbin{\circ}}

\newcommand\epsilontype{\boldsymbol{\varepsilon}}

\newcommand\mytheoremdefs{
\usepackage{theorem}
\theorembodyfont{\rm}
\newtheorem{definition}{Definition}[section]

}

\newcommand\lf{\mathbin\searrow}
\newcommand\rf{\mathbin\swarrow}

\newcommand\derives{\mathrel{\stackrel{*}{\Rightarrow}}}

\newcommand\lambek{\mathrel{\triangleleft}}
\newcommand\infern[3]{\begin{prooftree} #1 \justifies #2 \using (#3) \end{prooftree}}
\newcommand\infer[2]{\begin{prooftree} #1 \justifies #2 \end{prooftree}}
\newcommand\lamb{\;}

\newcommand\lam[1]{\lambda #1\mkern1mu . \mkern1mu}
\newcommand\sem[1]{[\mkern-2.5mu[ #1 ]\mkern-2.5mu]} 

\newcommand\lfl{\lf\textsc{l}}
\newcommand\lfr{\lf\textsc{r}}
\newcommand\rfl{\rf\textsc{l}}
\newcommand\rfr{\rf\textsc{r}}
\newcommand\lfe{\lf\textsc{e}}
\newcommand\rfe{\rf\textsc{e}}
\newcommand\leps{\epsilontype\textsc{L}}
\newcommand\reps{\epsilontype\textsc{R}}
\newcommand\prodl{\lamprod\textsc{l}}
\newcommand\prodr{\lamprod\textsc{r}}

\newcommand\wherefigure{}

\newcommand\myarray{\renewcommand\arraystretch{1.5}}

\newcommand\qed{}

\mytheoremdefs

\title{Command injection attacks, continuations, and the Lambek calculus}

\author{Hayo Thielecke
\institute{School of Computer Science\\University of Birmingham}
\email{H.Thielecke@cs.bham.ac.uk}
}

\def\titlerunning{Command injection attacks, continuations, and the Lambek calculus}
\def\authorrunning{H.~Thielecke}

\date{}

\begin{document}

\maketitle

\begin{abstract} 
This paper shows connections between command injection attacks, continuations, and the Lambek calculus: certain command injections, such as the tautology attack on SQL, are shown to be a form of control effect that can be typed using the Lambek calculus, generalizing the double-negation typing of continuations.
Lambek's syntactic calculus is a logic with two implicational connectives taking their arguments from the left and right, respectively. These connectives describe how strings interact with their left and right contexts when building up syntactic structures.  The calculus is a form of propositional logic without structural rules, and so a forerunner of substructural logics like Linear Logic  and Separation Logic.  
\end{abstract}

%
%
%
%
%
%
%
%


\section{Introduction}
\label{secintro}

The aim of this paper is to draw connections between three at first sight disparate topics ranging from the practical to the theoretical side of computer science:
\begin{enumerate}
\item Command injection attacks;
\item continuations and control effects; 
\item the Lambek calculus, a presentation of syntax as a logic or type system.
\end{enumerate}
Depending on the reader's background, the following may serve as an introduction to command injections, the Lambek calculus, or both. Continuations will serve as the glue between these topics, so to speak, and a basic familiarity with control operators and their typing is assumed.

We briefly recall some background on continuations.
Continuations in one form or another occur in many areas of computer science, ranging from compiling to logic. Like many fundamental concepts, they
have been discovered independently~\cite{reynoldsdisc}, and we may even see G\"odel's work  on double negation as one of the first such discoveries. 

Consider an expression language with a control operator \texttt{return}, as given in Figure~\ref{figsimplecont}. As shown in some of the earliest work on continuation semantics~\cite{stracheywadsworth}, 
such a language can be given a semantics by taking a continuation as a parameter.
\begin{figure}
\begin{eqnarray*}
E &::=& E + E
\\
& \mid & n
\\
& \mid & \texttt{return }n
\end{eqnarray*}
\begin{eqnarray*}
\sem{E_{1} + E_{2}} 
&=& 
\lam k \sem{E_{1}}(\lam{x_{1}} \sem{E_{2}} (\lam{x_{2}}k(x_{1} + x_{2}))
\\
\sem n & = & \lam k k\,n
\\
\sem{\texttt{return }n} & = & \lam k n
\end{eqnarray*}
\caption{Control operator \texttt{return} and its continuation semantics}
\label{figsimplecont}
\end{figure}

For example, the expression
\[
(\texttt{return }42) + 666
\]
evaluates to 42. Intuitively, this is because the evaluation context 
\[
(\hole + 666)
\]
has been discarded by the control operator.

The typing of the continuation semantics is a generalized double negation:
\[
\sem E : (\texttt{int} \to {\texttt{int}}  ) \to {\texttt{int}} 
\]

The typical presentation of continuation passing, following Plotkin~\cite{plotkincont}, uses  $\lambda$-calculus, but  
much of the machinery of continuations works in a more general situation. 
If fewer structural rules 
are assumed
(omitting Contraction and Weakening), then connections with Linear Logic emerge~\cite{LinUCHOSC,filinskilinear}. Lambek's syntactic calculus~\cite{lambekcalculus} goes further and removes the  Exchange rule as well, making it a canonical logic for reasoning about strings.
Control operators, by giving access to the current continuation, have an effect of the surrounding evaluation context. Analogously, in the Lambek calculus, a binary operator has a syntactic effect in the sense that it consumes some of its syntactic context, as given by symbols to its left and right.

\section{Command injection attacks}

In programming language theory, one usually assumes that all matters of parsing have been settled, so that the syntax is given as abstract syntax trees, rather than raw sequences of symbols,
before language features such as types or effects are considered.
However, complex software increasingly contains parsers and interpreters of various kinds, some for full programming languages, others for more restricted languages such as SQL or XML. 
Input to them is parsed at runtime, and may originate from untrusted sources.
Consequently, syntax 
becomes a problem again, impacting the safety and security of the interpreters.

Command injection attacks form a large class of attacks on software (for an overview, see texts on secure programming, such as Dowd et al~\cite{dowd-software}).
 They may happen whenever user-malleable and potentially malicious fragments in some syntax are spliced into a syntactic context such that the resulting string is parsed and interpreted. 
It is crucial for the attack that the fragments to be combined are raw text that still has to be parsed, rather than some structured format such as abstract syntax trees.
Of course an attacker could just inject syntactically invalid gibberish and provoke parsing errors. Depending on the robustness of error handling, that could amount to a mere nuisance or a denial of service attack. However,  command injection attacks are far more pernicious by creating strings that are successfully parsed and therefore interpreted. By gaining access to the interpreter to run code of their choosing, attackers can violate integrity and confidentiality, rather than merely triggering errors.

SQL command injection~\cite{classificationsql} attacks are perhaps the best known example;  in this case the constructed strings are SQL queries that are interpreted by the database management system.
In some variants~\cite{classificationsql} of SQL command injection attacks, the attacker relies on injecting code with side effects, such as a \texttt{DROP} statement in SQL that destructively updates the database. In this paper we will however concentrate on a class of attacks that do not require side effects in the injected code and rely purely on \emph{syntactically} subverting the constructed string. The so-called tautology attack is a notorious example. Simply put, a malicious user injects the string \texttt{OR 1 = 1}, which when combined with a Boolean test renders that test tautologically  true and hence useless.

Command injection attacks are by no means confined to SQL injections. They may arise whenever data is mixed with code in the broadest sense of the word; for instance, XPath injection attacks are a recent example. So even if SQL attacks are defeated by standard secure coding techniques, 
it is reasonable to expect that more vulnerabilities and attacks will emerge as dynamic scripting languages and XML/HTML based technologies proliferate, in which the mixing of code and data
or in-band signalling
that security experts warn against is a widespread risk.

 At first sight, command injection attacks appear to be due to side effects in the interpreted language. Indeed, some attacks rely on the presence of effectful operations, such as inserting \texttt{UPDATE} or \texttt{DROP} in  so-called ``piggyback''
SQL command injection attacks. The tautology attack, however, afflicts the purely functional language of Boolean expressions, and it does so by syntactic means rather than any side effects.  
The ``essence''~\cite{wassermanessence}  of such attacks can be made precise in terms of parse trees. 
Intuitively, the programmer has an implicit understanding that the data supplied by the user should be slotted into the parse tree of the query as a leaf, below the comparison to \texttt{password}. Instead, the insertion of the operator \texttt{OR} rearranges the parse tree, so that the operator is above the test, rendering it ineffective by disjunction with the tautology. In this paper, we will focus exclusively on such
syntactic attacks on purely functional languages.

A simple example of  a syntactic command injection attack is known as a tautology attack. 
Suppose a dynamically constructed SQL query contains a comparison of the string \texttt{password} to some string supplied by the user, in order to check the user's authorization. Details of SQL syntax are not important here, but the idea of the syntactically malicious input is as follows.
The query is constructed by concatenating a  string
ending in 
``\texttt{password = '}'' with the user input to construct a boolean expression. This test is part of an SQL statement, as in`` \texttt{SELECT * FROM table WHERE}\ldots''.
 If the user supplies the input ``\texttt{foo}'', the concatenation contains the test ``\texttt{password = 'foo'}'', as intended. In an attack, the user injects an operator, by supplying the input  ``\texttt{foo' OR 1 = 1 --}''. The resulting test is
 \[
 \texttt{password = 'foo' OR 1 = 1}
 \]
  which always evaluates to true due to the tautology \texttt{1 = 1}. Using this technique, attackers may read confidential data for other users, bypass password authentication, and the like, with many variations on this theme of injecting operators~\cite{classificationsql}.

Given that the phenomenon of syntactic command injection is so general, 
and independent of many details of the particular technologies being exploited, 
we aim to address it at the appropriate level of abstraction. We would like to  reason about the syntactic effect, as it were, that a malicious input has on its context, similar to the way that a type and effect system~\cite{polyeffects} lets us reason about side effects. The effect can be subtle, as
a malicious input string needs to conform closely to the structure of  the surrounding string it is intended to attack. If for instance some delimiters are added to the latter, the original attack string may fail and produce only something syntactically ill formed.

\begin{figure}
\begin{description}
\item[String with a hole:] \texttt{password = }$\hole$

\item[Legitimate input:]  \texttt{foo}

\item[Combined string:]  \texttt{password = foo}

\item[Malicious input:] \texttt{foo OR 1 = 1}

\item[Combined string:] \texttt{password = foo OR 1 = 1}

\end{description}
\caption{Tautology attack}
\label{figtautattack}
\end{figure}

 A central thesis of this paper is that the required logical tools are already available in mathematical linguistics---perhaps surprisingly so. Lambek's syntactic calculus~\cite{lambekcalculus}
 describes syntax with two (left and right) function types that capture how a phrase takes other phrases as arguments from the left or right. Using these connectives, we will explain how the harmless input
 ``\texttt{foo}'' differs from the malicious input ``\texttt{foo OR 1 = 1}''  that can take its place. In essence, the malicious input is effectful in that it seizes part of its context, just as a control operator does with its evaluation context. See Figure~\ref{figtautattack}  (quotes are elided for simplicity).
%

%

\section{Syntax and the Lambek calculus}
\label{seccalculus}

As context-free grammars and parser generators for them are universally used in computer science, we will assume that the language we wish to reason about is given by an unambiguous context-free grammar. We will then  use
the Lambek calculus on top of the given grammar, not to define the language, but to describe the way its phrases combine.


We recall that a context-free grammar $\mathbf G = (\mathbf T, \mathbf N, \mathbf P, S)$ consists of a finite set $\mathbf T$ of terminal symbols, a finite set $\mathbf N$ of non-terminal symbols, a start symbol $S\in \mathbf N$, and a finite relation $\mathbf P \subseteq \mathbf N \times (\mathbf N \cup \mathbf T)^{*}$ relating non-terminal symbols to strings of symbols. The elements $(A,\alpha)$ of $\mathbf P$ are called the rules or productions of the grammar, and often written as $A\ ::= \alpha$. (We avoid the common notation 
$A\to\alpha$, as it clashes with that for function types.) 

We follow some notational conventions for grammars~\cite{dragonbook}. 
We write terminal symbols in typewriter font, as in ``\texttt a'' and ``\texttt =''.
Non-terminal symbols are ranged over by $A$, $B$, $C$, while $X$ and $Y$ may be a terminal or a non-terminal symbol. Sentential forms (strings that may contain both non-terminals and terminals) are written as $\alpha$, $\beta$, $\gamma$ and $\delta$. Words (strings of only terminal symbols) are ranged over by $w$, $v$, $u$.
The empty sequence is written as $\varepsilon$. 
The one-step derivation relation $\Rightarrow$ holds between any two strings of the form
\[
\beta\,A\,\gamma \Rightarrow \beta\,\alpha\,\gamma
\]
whenever there is a production $(A,\alpha)\in \mathbf P$. The reflexive transitive closure of $\Rightarrow$ is written as $\derives$.

A grammar is called unambiguous if there is no word $w$ that has two different parse trees with root $S$.
If we assume our grammar to be unambiguous, we are justified in speaking of ``the'' parse tree of a word.
For a non-terminal symbol
$A$, we say $A$ is useless if it does not participate in the derivation of any words, that is, if there are no $\alpha$, $\beta$ and $w$ such that
\[
S \derives \alpha\,A\,\beta \derives w
\]
We will assume that there are no useless non-terminals in the grammar (as deleting them will not change the language of the grammar). If the grammar is unambiguous and contains no useless symbols, the language of each non-terminal is also unambiguous. Unambiguous grammars are important in practice because compilers and interpreters compute meanings by induction over the parse tree; if there could be more than one such tree for a given input, there might be unintended outcomes.


  When first reading about Lambek's syntactic calculus, one may perhaps be puzzled about whether to conceive of it as a form of syntax, a type system, or a logic. It is in a sense all of these, and that flexibility may be an advantage. There are two equivalent presentations of the calculus: the first as subtyping (to use current terminology), the other as a propositional logic in the style of Gentzen's sequent calculus.
  
Before going into the formal definitions of the calculus, it may be helpful to provide some intuition about its intended meaning, particularly compared to context-free grammars. Suppose we want to express that the operator \texttt{OR} takes a truth value $T$  from the left and right, respectively,  and produces a truth value. Using context-free grammars, we could write a grammar rule like the following:
\[
T \gramrule T \; \texttt{OR} \;T
\]
(To keep the discussion simple, let us ignore the problem of ambiguous grammars for the moment.)
In the Lambek calculus, we would express the same syntactic situation differently. We would say that there is a type of of words that produce a $T$ if a $T$ is placed to the left of them, which we write as $T\lf T$. Moreover, there is a type of words that produce the latter type if another $T$ is placed to the right of them, which is written as $ (T \lf T) \rf T$. That gives us a type of binary operators expecting a $T$ on either side. Stating that \texttt{OR} is such a binary operator amounts to a subtyping judgement  for the type \texttt{OR} (which contains exactly the word \texttt{OR}):
\[
\texttt{OR} \subtype (T \lf T) \rf T
\]
A useful intuition to bear in mind when reading the syntactic calculus is that the left-hand side is meant to be a subset of the right-hand side. In our example here, the set containing only \texttt{OR} is a subset of the set of binary operators, but not necessarily conversely, as there may be other such operators. Note that the order of writing is reversed compared to grammar rules: a grammar rule
$A \gramrule B$ corresponds to $B \subtype A$.

If we also have $\texttt{1 = 1} \subtype T$, then we see that the partial application of \texttt{OR} to it still expects a $T$ on its left:
\[
\texttt{OR 1 = 1} \subtype T \lf T
\]
Thus we can construct various operators by partial application (currying), as is familiar from functional programming. It would be possible to capture the syntax of a language entirely with
such judgements, without the need for a context-free grammar. However, in our setting we assume a fixed grammar is given, and we use the Lambek calculus for reasoning about fragments of words like the \texttt{OR 1 = 1}  above.

\begin{figure}\wherefigure  
\[
\renewcommand\arraystretch{4}
\begin{array}{c@{\hspace{4em}}c}
   \infer{\varphi_{1} \lamprod \varphi_{2} \subtype \psi}{\varphi_{2} \subtype \varphi_{1} \lf \psi}
&
   \infer{\varphi_{1} \lamprod \varphi_{2} \subtype \psi}{\varphi_{1} \subtype  \psi \rf \varphi_{2}}
\\
  \infer{\varphi_{2} \subtype \varphi_{1} \lf \psi}{\varphi_{1} \lamprod \varphi_{2} \subtype \psi}
  &
     \infer{\varphi_{1} \subtype  \psi \rf \varphi_{2}}{\varphi_{1} \lamprod \varphi_{2} \subtype \psi}
  \\
   \infer{}{\varphi \subtype \varphi}
 &
  \infer{\varphi_{1} \subtype \varphi_{2} \qquad \varphi_{2} \subtype \varphi_{3}
  }{\varphi_{1} \subtype \varphi_{3}}
\\
\end{array}
\]
  \[
\renewcommand\arraystretch{3.5}
\begin{array}{c}
  \infer{}{\varphi_{1} \lamprod (\varphi_{2} \lamprod \varphi_{3}) \subtype (\varphi_{1} \lamprod \varphi_{2})
  \lamprod \varphi_{3}}
  \\  
  \infer{}{(\varphi_{1} \lamprod \varphi_{2}) \lamprod \varphi_{3} \subtype \varphi_{1} \lamprod (\varphi_{2}
  \lamprod \varphi_{3})}
  \end{array}
\]
\caption{Lambek's syntactic calculus, subtyping version}
\label{figsyntactic}
\vspace{.5ex}
\hrule
\end{figure}

\begin{figure}\wherefigure  
\[
\infer{(A,X_{1}\ldots X_{n}) \in\mathbf P}{X_{1}\lamprod \ldots \lamprod X_{n} \subtype A}
\]
\[
\renewcommand\arraystretch{3}
\begin{array}{c@{\hspace{4em}}c}
  \infer{}{\epsilontype \lamprod \varphi \subtype \varphi}
  &
    \infer{}{\varphi \lamprod \epsilontype \subtype \varphi}
  \\
    \infer{}{\varphi \subtype \epsilontype \lamprod \varphi}
      &
    \infer{}{\varphi \subtype \varphi \lamprod \epsilontype }
  \end{array}
\]
\caption{Additional rules for subtyping}
\label{figmoresubtype}
\vspace{.5ex}
\hrule
\end{figure}


We assume that a fixed context-free grammar 
 \[
 \mathbf G = (\mathbf T, \mathbf N, \mathbf P, S)
 \]
of interest is given, and we define a version of the syntactic calculus
specific to that grammar by using its symbols as the base types and importing its rules as axioms.
\begin{definition}
\label{defsyntactictypes}
\normalfont
The types of (our variant of) the Lambek calculus are built up 
from the (terminal or non-terminal) symbols of our context-free grammar (ranged over by $X$)
using the left and right arrow connectives as well as the product connective.
\[
\myarray
\begin{array}{rcl@{\qquad}l}
\varphi, \psi,\pi &::=& X &\textrm{(Grammar symbol in $\mathbf N\cup\mathbf T$)}
\\
&\mid& \varphi\lf \psi & \textrm{(Left implication)}	
\\
&\mid& \psi\rf \varphi & \textrm{(Right implication})	
\\
&\mid& \varphi \lamprod \psi & \textrm{(Product)}	
\\
&\mid&  \epsilontype & \textrm{(Empty string type)}
\end{array}
\]
\end{definition}

\begin{definition}[Syntactic calculus, subtyping variant]
\label{defsyntactic}
The syntactic calculus consists of subtyping judgements of the form
\[
\varphi \subtype \psi
\]
where $\varphi$ and $\psi$ are defined as in Definition~\ref{defsyntactictypes}.
The rules for $\subtype$ are given in Figures~\ref{figsyntactic} and~\ref{figmoresubtype}.
\end{definition}

In the literature, the two implications are written as forward and backward slashes, ``/'' and ``\textbackslash''. Reading
such formulas can be tricky, particularly since two conventions exist. Lambek's notation places the result on top 
and reflects whether parameters are taken from left or right; Steedman's notation instead emphasizes the directionality
of functions by placing the parameter on the left and the result on the right.
We follow the Lambek style, but add arrowheads, writing ``$\lf$'' and ``$\rf$'', to make it easier to see where the parameter and where the result is.

  For reading nested implications, it is useful to bear in mind whether the arrows are pointing inward or outward. The following two types are isomorphic:
  \[
   ( \varphi_{1} \lf \psi ) \rf \varphi_{2}
  \mbox{ and }
  \varphi_{1} \lf (\psi \rf \varphi_{2} )
  \]
  Intuitively, it makes no difference whether a binary operator consumes its left operand $\varphi_{1}$ or its right operand $\varphi_{2 }$ first.  We may write $ \varphi_{1} \lf \psi  \rf \varphi_{2}$
  to mean either of them, just as brackets can be omitted due to  $\lamprod$ being associative. By contrast, the 
  two types where $\psi$ occurs in a doubly negative position, as in: 
   \[
   ( \varphi_{1} \rf \psi )  \lf \varphi_{2} 
  \mbox{ and }
 \varphi_{2} \rf (\psi \lf \varphi_{1} )
   \] 
are genuinely different, even when $\varphi_{1} = \varphi_{2}$. Such doubly-negated types will
be pertinent later on, particularly in Section~\ref{secciaeffects}. 

The rules of the syntactic calculus are divided into those that are taken directly from Lambek's paper~\cite{lambekcalculus}, gathered in Figure~\ref{figsyntactic}, and additional rules we add in this paper
for  using of the calculus on top of  a fixed context-free grammar, presented in Figure~\ref{figmoresubtype}.
 
In logical terms, the four rules for the implications in Figure~\ref{figsyntactic} are quite natural if one thinks of implications (or arrow types) as
adjoints of conjunctions (or products).
In linear logic, the linear implication
$\multimap$ is adjoint to $\varphi \otimes (-)$. In separation logic, the separating implication
$\wand$ is adjoint to the separating conjunction $\varphi * (-)$.
 In the Lambek calculus,  the product is not commutative, so that
$(-) \lamprod \varphi$ and $\varphi \lamprod (-)$ are not interchangeable. 
Consequently,  there are two different adjoints $\lf$ and $\rf$.
The other four rules state that
 the subtyping relation $\subtype$ is reflexive and transitive, and that the product $\lamprod$ is associative.

The rules in  Figure~\ref{figsyntactic} are the logical core of the calculus that applies to any language. In order to specialize
the calculus to a particular language, we need additional axioms.
In our case here, we import all productions of the  given context-free grammar into the subtyping relation by adding axiom schemas stating that
the product of the symbols on the right-hand side of the production is a subtype of the non-terminal symbol on the left of the production. Note that the order of the subtyping is the reverse of the way grammars are written; it is in reduction rather than derivation order. 
As we can have epsilon productions (having an  empty string on the right-hand side) in the grammar, we need to represent the empty string $\varepsilon$ in the syntactic calculus as well. We do so by adding a type constant called
$\epsilontype $ and rules making it a left and right unit for product.
Logically, $\epsilontype$ is a natural addition to the calculus, in that it is the nullary analogue of Lambek's binary $\lamprod$ connective. These rules are given in Figure~\ref{figmoresubtype}.

\begin{figure}\wherefigure 
\[
\renewcommand\arraystretch{4}
\begin{array}{c@{\hspace{4em}}c}
\infern{\Phi \lambek \varphi \qquad \Psi\lamb \psi\lamb \Pi
\lambek\pi }{\Psi\lamb(\psi\rf\varphi)\lamb\Phi\lamb\Pi\lambek \pi}{\rfl}
&
\infern{\Phi\lamb \varphi \lambek \psi}{\Phi \lambek \psi\rf \varphi}{\rfr}
\\
\infern{\Phi \lambek \varphi \qquad \Psi\lamb \psi\lamb \Pi
\lambek\pi }{\Psi\lamb\Phi \lamb(\varphi\lf\psi)\lamb\Pi\lambek \pi}{\lfl}
&
\infern{\varphi\lamb \Phi \lambek \psi}{\Phi \lambek \varphi \lf \psi}{\lfr}
\\
\infern{\Phi \lamb \varphi \lamb \psi \lamb \Psi \lambek \pi}{\Phi \lamb (\varphi \lamprod \psi) \lamb \Psi \lambek \pi}{\prodl}
&
\infern{\Phi \lambek \varphi\qquad \Psi \lambek \psi}{\Phi\lamb\Psi \lambek \varphi \lamprod \psi}{\prodr}
\\
\infern{\Phi \lambek \varphi\qquad \Psi\lamb \varphi \lamb \Pi \lambek\psi }{\Psi\lamb \Phi\lamb \Pi\lambek \psi}{\textsc{cut}}
&
\infern{}{\varphi \lambek \varphi}{\textsc{ax}}
\end{array}
\]
\caption{Sequent calculus variant of Lambek's syntactic calculus}
\label{figlambek}
\vspace{1ex}
\hrule
\end{figure}

\begin{figure}\wherefigure  
\[
\infern{(A,\alpha) \in\mathbf P}{\alpha \lambek A}{\mathbf P{\lambek}}
\]
\vspace{1em}
\[
\infern{\Phi\lamb\Psi \lambek \varphi}{\Phi\lamb \epsilontype \lamb \Psi \lambek \varphi}{\leps}
\hspace{4em}
\infern{}{\quad \lambek \epsilontype}{\reps}
\]
\caption{Additional rules for sequents}
\label{figmoresequent}
\vspace{.5ex}
\hrule
\end{figure}

Lambek~\cite{lambekcalculus} also defines a sequent calculus, as this yields a decision procedure.
In the literature, this sequent presentation is often referred to simply as the Lambek calculus.
The calculus has left and right rules for the connectives, and it lacks all structural rules, going even further than Linear Logic and Separation Logic by banishing the Exchange rule~\cite{vanbenthemaction}. Hence it distinguishes between a left and a right implication connective.

\begin{definition}[Sequent presentation of the calculus]
Let $\varphi$, $\psi$ and $\pi$ range over types as in Definition~\ref{defsyntactictypes}.
We let the capital Greek letters $\Phi,\Psi$ and $\Pi$ range over sequences of the form $\varphi_{1}\ldots \varphi_{n}$, written without separating commas. 
 Judgements 
 are of the form $\Phi\lambek\varphi$, using the inference rules  listed in Figures~\ref{figlambek} and~\ref{figmoresequent}. 
\end{definition}

The Lambek calculus has a simple denotational semantics. In particular, 
the two implications are interpreted as left and right language difference, product as concatenation, and
judgements are interpreted as 
language inclusion~\cite{vanbenthemaction}.
For our version of the calculus, built on top of a context-free grammar, its semantics is as follows:

\begin{definition}
\label{deflambeksem}
\normalfont
The denotation of a type in the syntactic calculus is a set of words, defined inductively as follows: 
\myarray
\begin{eqnarray*}
\sem{X} & = & \{w\in \mathbf T^{*} \mid X \derives w \}
\\
\sem{\varphi\lf\psi} &=& \{w\in \mathbf T^{*}  \mid \forall v\in \mathbf T^{*}.v\in \sem{\varphi} \textrm{ implies }
v\,w \in \sem\psi\}   
\\
\sem{\psi\rf\varphi} &=& \{w\in \mathbf T^{*}  \mid \forall v\in \mathbf T^{*}.v\in \sem{\varphi}  \textrm{ implies }
w\,v\in \sem\psi\}  
\\
\sem{\varphi_{1}\lamprod\varphi_{2}}  &=& \{w_{1}\,w_{2 }\in \mathbf T^{*} \mid w_{1} \in\sem{\varphi_{1}} \mbox{ and } w_{2} \in\sem{\varphi_{2}}
\} 
\\
\sem{\epsilontype} &=& \{\varepsilon \}
\end{eqnarray*}
 The semantics of a logical context $\Phi = \varphi_{1} \ldots \varphi_{n}$ is the same as that of the $n$-fold product of $\varphi_{j}$, unless the sequence is empty, in which case it is the same as $\epsilontype$:
\begin{eqnarray*}
 \sem{\Phi} &=& \{\varepsilon\} \mbox{ if $\Phi$ is the empty context}
\\
 \sem{\varphi_{1}\lamb \ldots \lamb \varphi_{n}} 
&=& 
\{ w \in \mathbf T^{*} \mid w= w_{1}\ldots w_{n} \mbox{ where }
\\&&
\hphantom{\{}
 w_{1}\in \sem{\varphi_{1}}, \ldots, w_{n}\in \sem{\varphi_{n}} \,\}
\end{eqnarray*}
\end{definition}


\section{Reasoning about syntactic effects}
\label{secciaeffects}

\begin{figure*}
\begin{center}
\begin{tikzpicture}
[level distance = 5ex]
\tikzstyle{level 1}=[sibling distance=18ex]
\tikzstyle{level 2}=[sibling distance=9ex]
\tikzstyle{level 3}=[sibling distance=6ex]
\tikzstyle{level 4}=[sibling distance=3ex]
\path (0,0)
node  {$E$}
child
{
node  {$C$}
child {
node  {$T$}
child {node {$V$} 
child {node{\texttt a}}
}
child {node{\texttt =}}
child {node {$V$} 
child {node{\texttt b}}
}
}
child {node  {$D$} child {node{$\varepsilon$}}}
}
child {node {$F$}
child { node{\texttt{OR}}}
child {
node  {$C$}
child {
node  {$T$}
child {node {$V$} 
child {node{\texttt 1}}
}
child {node{\texttt =}}
child {node {$V$} 
child {node{\texttt 1}}
}
}
child {node  {$D$} child {node{$\varepsilon$}}}
}
child {node {$F$}
child { node{$\varepsilon$}}}
}
;
[level distance = 5ex]
\tikzstyle{level 1}=[sibling distance=10ex]
\tikzstyle{level 2}=[sibling distance=8ex]
\tikzstyle{level 3}=[sibling distance=4ex]
\path (7,0)
node  {$E$}
child
{
node  {$C$}
child
{
node  {$T$}
child {node {$V$} 
child {node{\texttt a}}
}
child {node{\texttt =}}
child {node {$\hole$}}
}
child {node  {$D$} child {node{$\varepsilon$}}}
}
child {node  {$F$} child {node{$\varepsilon$}}}
;
\end{tikzpicture}
\end{center}
\caption{Parse tree for \texttt{a = b OR 1 = 1} and partial parse tree for \texttt{a = }$\hole$}
\label{figureparsetree}
\end{figure*}

In this section, we first investigate a command injection attack as an example of reasoning in the syntactic calculus. Building on what can be gleaned from that example, we then place it into a wider context of types and effects.

\label{subsectaut}

We define a toy grammar of Boolean expressions that is sufficient for discussing tautology attacks.
The grammar uses a standard technique to avoid ambiguity and to ensure that conjunction binds more tightly than disjunction~\cite{dragonbook}.
An expression $E$ is a disjunction of conjunctions $C$ of
equality tests $T$ between values $V$.
A series of applications of \texttt{AND} is parsed as a $C$, but no \texttt{OR} can appear in a $C$.

\begin{eqnarray*}
E &::= & C\;F 
\\
F &::=& \texttt{OR}\;C\;F
\\
&\mid& \varepsilon
\\
C &::=& T\,D
\\
D &::=& \texttt{AND}\;T\,D
\\
&\mid& \varepsilon
\\
T &::=& V\; \texttt{=}\; V 
\\
V &::=& 1 \mid \ldots \mid \texttt a \mid \texttt{b} \mid \ldots
\end{eqnarray*}

We will now reason about the interaction between malicious inputs and vulnerable contexts using the syntactic calculus in its logical variant. 
The presentation as a sequent calculus, with left and right rules for the connectives, may look unfamiliar compared to type systems, which are usually in natural deduction style. 
Nonetheless, elimination rules for the arrow types are derivable:
\[
\begin{prooftree}
\Phi \lambek \varphi \qquad \Psi \lambek\varphi \lf \psi
\justifies
\Phi\lamb \Psi\lambek \psi
\using{(\lfe)}
\end{prooftree}
\]
A symmetric elimination rule $(\rfe)$ exists for $\rf$.
The proof for deriving $(\lfe)$ is as follows:
\[
\begin{prooftree}
\Psi \lambek\varphi \lf \psi
\qquad 
\[
\Phi \lambek \varphi \qquad 
\[
\justifies
\psi\lambek \psi
\using{(\textsc{ax})}
\]
\justifies
\Phi \lamb (\varphi\lf\psi) \lambek \psi
\using{(\lfl)}
\]
\justifies
\Phi\lamb \Psi\lambek \psi
\using{(\textsc{cut})}
\end{prooftree}
\]

Now suppose we have some code in which a string variable is concatenated with the string constant ``\texttt{a = }''. 
 The judgement  $\texttt{a =}\lambek T\rf V$
 tells us that the
  incomplete test expects a value to its right. If we supply such a value, say \texttt b, we infer using the derived elimination rule:
  \[
\infern{
\texttt{a}\,\texttt{=}\lambek T \rf V
\qquad
 \texttt{b} \lambek V}
 { \texttt{a = b} \lambek T}{\rfe}
\]

Now consider the attack string \texttt{b OR 1 = 1}. The essential point is that the attack reverses the role of operator and operand when concatenated with the fragment \texttt{a =}. In our calculus that is captured by the judgement
 \[
 \texttt{b OR 1 = 1} \lambek ( T \rf V)\lf E
 \]
 To infer this, we first note that we can derive in the grammar
 \[
 E \derives T\;\texttt{OR 1 = 1}
 \]
 which implies $T\;\texttt{OR 1 = 1}\lambek E$. 
 From that, we construct the following proof:
\[
\begin{prooftree}
\[
\[
\[
(V, \texttt{b}) \in \mathbf P
\justifies
\texttt{b} \lambek V
\using{(\mathbf P\lambek)}
\]
\quad
\[
\justifies
T \lambek T
\using{(\textsc{ax})}
\]
\justifies
(T \rf V) \lamb \texttt{b}
 \lambek T
 \using{(\rfl)}
 \]
\quad
T \lamb\texttt{OR 1 = 1} \lambek E
\justifies
(T \rf V) \lamb\texttt{b OR 1 = 1} \lambek E
\using{(\textsc{cut})}
\]
\justifies
\texttt{b OR 1 = 1} \lambek (T \rf V) \lf E
\using{(\lfr)}
\end{prooftree}
\]

The two syntax fragments fit together to build an expression:
\[
\infern{
\texttt{a}\,\texttt{=}\lambek T \rf V
\qquad
 \texttt{b OR 1 = 1} \lambek ( T \rf V)\lf E}
 { \texttt{a = b OR 1 = 1} \lambek E}{\lfe}
\]
The fragment \texttt{a =} is now in the operand position of the application, rather than the operator position it had in $\texttt{a = b}\lambek T$.

As Figure~\ref{figureparsetree} shows, the parse tree for 
 \texttt{a = b OR 1 = 1} does not arise from completing the partial parse tree for \texttt{a = }$\hole$ displayed to its right, where $\hole$ indicates the ``hole'' position in the partial parse tree and syntax fragment.
 
The common name in the software security literature is Tautology attack, as it is the tautology that renders the test trivially true. However, in terms of reshaping the parse tree, 
the crucial ingredient is the
 fact that the injected operator \texttt{OR} has a lower precedence that the adjacent operator \texttt =, as the low precedence causes the \texttt{OR} node to move up in the parse tree.
  
Whether or not this way of combining pieces of syntax is an attack or a useful way to build up strings 
depends on what type we consider the function to have. 

\section{Double negation in command injection and linguistics}
\label{subsecorderneg}

 Note that the string with the syntactic effect is very sensitive to the context on which it has an effect. 
If we merely change the order in the latter, replacing $\texttt{a = }\hole$ with $\hole \texttt{ = a}$, the original attack does not work anymore, producing only an ungrammatical  string. 
(In software security practice, that means attackers may need some reverse engineering skills to craft malicious input that fits into the syntactic context like a key into a lock.)
The two connectives $\lf$ and $\rf$ capture such ordering accurately. For injecting  into $\hole\texttt{ = a}$, the 
attack string is symmetric to the one above, with all implications  reversed:

   \begin{center}
\myarray
\begin{tabular}{lll}
String & has type & fitting into context
\\
  \texttt{b OR 1 = 1} & $( T \rf V)\lf E$
  & $\texttt{a = }\hole$
 \\
 \texttt{1 = 1 OR b} & $E \rf (V \lf T)$ & $\hole\texttt{ = a}$
 \\
\end{tabular}
\end{center}


The main use of the Lambek calculus and related formalisms such as categorial grammar has been in linguistics rather than computer science (although Lambek's original paper discusses examples from logic along with those from natural language). Nonetheless, there are some intriguing parallels to the situations we have discussed.

 Consider a naive syntax for English sentences. 
We have a type \senttype{} of sentences and a type \textbf{Noun} of nouns. Names like
``\texttt{Alice}'' and ``\texttt{Bob}''
have type \textbf{Noun}. In the syntactic calculus, a transitive verb has a type like a binary operator, for instance 
  \[
\texttt{knows}
\lambek
\textbf{Noun} \lf (\senttype \rf \textbf{Noun})
\]
So we can derive sentences like`` \texttt{Alice knows Bob}'' in the same way as deriving Boolean expressions like  \texttt{a = b OR 1 = 1}. 
Lambek~\cite{lambekcalculus} observes that pronouns like 
\texttt{he}
and 
\texttt{him}
may occur in some positions in which nouns may occur. However, pronouns are  more sensitive to their position,
because`` \texttt{he}'' has to occur to the left of the verb, whereas ``\texttt{him}'' 
needs to be 
to the right of the verb.
The calculus captures this grammatical fact by giving the two different double negations as the types of ``\texttt{he}''
and 
``\texttt{him}'':
        \begin{center}
\myarray
\begin{tabular}{lll}
String & has type & fitting into context
\\
\texttt{he} &
$\senttype \rf (\textbf{Noun} \lf \senttype )$
&
$\hole\texttt{ knows Alice}$
 \\
\texttt{him}  
& $( \senttype \rf \textbf{Noun}) \lf \senttype$ 
&
  $\texttt{Alice knows }\hole$
 \\
\end{tabular}
\end{center}

 Compare the difference between
 injection to the left or the right of the equality test discussed  above.

\section{Syntactic effects and  control effects}
\label{subseccontrol}

Rather than supplying the expected type $V$, the attack string supplies a kind of generalized double negation of $V$, or more precisely, a $V$ inside the negative position of two implications, as in 
  \[
   ( \varphi_{1} \rf V )  \lf \varphi_{2} 
  \mbox{ and }
 \varphi_{2} \rf (V \lf \varphi_{1} )
   \] 
This typing generalizes the double negation of a formula $A$ in logic, namely
\[
(A \to \bot)\to\bot
\] 
The raising to a doubly-negated type  is reminiscent of control operators in programming languages, and specifically the way that continuation passing style (CPS) introduces a form of double negation. 

As a brief reminder of control operators, we
consider the following simple use of the control operator \texttt{call/cc}: 
\[
(\texttt{call/cc}(\lam k{42+(k\,2})))+1
\]
Operationally, the current continuation is bound to the variable $k$ when the call to \texttt{call/cc} is evaluated. Continuations can be represented as  evaluation contexts~\cite{felleisenreasoning}, written as  terms with a hole. In our example, the continuation bound to $k$ could be written as
\[
\hole +1
\]
where $\hole$ stand for the hole.
When $k$ is invoked in the subexpression $(k\,2)$, the value 2 is plugged into the hole of the continuation, and the whole expression thereby evaluates to 2+1 = 3.
If the operational semantics of control operators is formalized in terms of evaluation contexts~\cite{felleisenreasoning}, a salient
 feature 
  is that their evaluation can move upward in the surrounding evaluation context. Compare how
   in Figure~\ref{figureparsetree}
   the  injected operator \texttt{OR} moves upward in the parse tree from where it was inserted by, so to speak, 
  elbowing itself across the node labelled  $T$.
  
The application $(k\,2)$ appears to be of type  \texttt{int}, in that it can be used as an argument of the operator $+$. The surrounding context, expecting an integer to be supplied, can be thought of  as a
function from $\texttt{int}$ to some answer type $\texttt{Ans}$. If a value occurs in the context, it is passed to the function, yielding an answer. However, if the expression inside the context has control effects, it does not simply supply a value to its context. Instead, it takes the context as an argument and manipulates it (in the example above, by discarding it and using the continuation bound to $k$ instead). Hence an expression with control effects of direct-style type \texttt{int} has a continuation-passing type 
that is a double negation of \texttt{int}:
\[
(\texttt{int} \to\texttt{Ans})\to\texttt{Ans}
\]
In programming language semantics, these double negations are inserted by continuation passing style transforms~\cite{plotkincont}. The resulting connection~\cite{griffin} to classical logic has been studied intensely. 
As a further refinement of this typing of control effects, an effect system can constrain how far up in the context the effect may reach~\cite{polyeffects,controleffects,effectspopl03}. 
In an effect system, we can control how effectful the argument of a function is. Suppose a function $f:C \to B$ is intended to  be pure, which means it has no effect. The function type for pure functions is written as 
$C \stackrel{\emptyset}{\longrightarrow} B$.
However, if $f$  calls a function passed as its argument,
that function also needs to be pure. In the effect system, we can express this by giving a type of this form:
\[
f: (A \stackrel{\emptyset}{\longrightarrow} B) \stackrel{\emptyset}{\longrightarrow} B
\]

In an effect system, one often has a notion of sub-effecting, where a function that has fewer latent effects can be used where one with potentially more effects is expected. This fits well with our view here that
a word with type $\varphi$ also has the two double negations of $\varphi$ as its type, but not conversely. 
  
To sum up, we would like to draw the following analogy between an expression with control effects and a syntactic command injection attack string: 
\begin{center}
\myarray
\begin{tabular}{lll}
Expression & Context expects & CPS type
\\
$(k\; 2)$ & \texttt{int} & $(\texttt{int} \to\texttt{Ans})\to\texttt{Ans}$
\\
\texttt{b OR 1 = 1} & $V$ & $( T \rf V)\lf E$
\\
 \texttt{1 = 1 OR b} &$V$&  $E \rf (V \lf T)$
\\
\end{tabular}
\end{center}

 Whereas the transformation of lambda terms into continuation passing style introduces nests of additional lambda abstractions, its analogue in the Lambek calculus is silent, so to speak. If a word $w$  expects some word $v$ on its right, we can regard $v$  as expecting such a $w$ on its left. So if $v$ is a $\varphi$ and $w$ a $\psi\rf \varphi$, then we can equally regard the same word $v$ as a $(\psi\rf \varphi)\lf\psi$.

More formally, there are two derivable rules for introducing  double negation:
\[
\begin{prooftree}
\Phi\lambek \varphi
\justifies
\Phi\lambek(\psi\rf\varphi)\lf \psi
\using{(\textsc{dnil})}
\end{prooftree}
\hspace{4em}
\begin{prooftree}
\Phi\lambek \varphi
\justifies
\Phi\lambek \psi\rf(\varphi\lf\psi)
\using{(\textsc{dnir)}}
\end{prooftree}
\]
These rules are derivable as follows:
\[
\begin{prooftree}
\[
\Phi \lambek \varphi \qquad 
\[
\justifies
\psi \lambek \psi
\using{\textsc{(ax)}}
\]
\justifies
 (\psi\rf\varphi) \lamb \Phi \lambek \psi
\using{(\rfl)}
\]
\justifies
\Phi\lambek(\psi\rf\varphi)\lf \psi
\using{(\lfr)}
\end{prooftree}
\hspace{5em}
\begin{prooftree}
\[
\Phi \lambek \varphi \qquad 
\[
\justifies
\psi \lambek \psi
\using{\textsc{(ax)}}
\]
\justifies
\Phi \lamb (\varphi\lf\psi) \lambek \psi
\using{(\lfl)}
\]
\justifies
\Phi\lambek  \psi \rf(\varphi\lf\psi)
\using{(\rfr)}
\end{prooftree}
\]

We recognize the syntactic control effects in the Lambek calculus as a form of continuation passing that goes even further 
in banishing structural rules
than
linear continuations~\cite{filinskilinear}
or linearly used continuations~\cite{LinUCHOSC}.

It is instructive to compare and contrast the two double-negation introductions in the Lambek calculus with double-negation introduction in intuitionistic and linear logic. Let us consider linear logic (as we can move from linear to intuitionistic logic by adding the Weakening and Contraction rules).
There is no distinction between left and right implications, with only a single introduction and a single elimination rule for the linear implication $\multimap$: 
\[
\infern{\Gamma,A\vdash B}{\Gamma \vdash A\multimap B}{\multimap I}
\qquad
\infern{\Gamma\vdash A\multimap B\qquad \Delta\vdash A}{\Gamma,\Delta \vdash B}{\multimap E}
\]
These rules give rise to a double negation introduction. Its proof relies on the ability to exchange formulas in the context: 
\[
\begin{prooftree}
\[
\[
 \[
  \justifies 
 A \multimap R 
 \vdash
 A \multimap R
 \]
 \qquad 
 \Gamma \vdash A
\justifies
A \multimap R,\Gamma \vdash R
\using{(\multimap E)}
\]
\justifies
\Gamma, A \multimap R \vdash R
\using{\textsc{(Exchange)}}
\]
\justifies
\Gamma \vdash (A \multimap R) \multimap R
\using{(\multimap I)}
\end{prooftree}
\]
The corresponding proof term is $\lam k{k\,x}$, or more precisely:
\[
x:A \vdash \lam k{k\,x} : (A \multimap R) \multimap R
\]
In the Lambek calculus, by contrast, there is no need for $\lambda$-abstraction and application. The continuation passing version of a word $w$ is just $w$ itself.

\section{Conclusions}
\label{secconcl}

%

In computer science generally, the Lambek calculus, particularly when presented as sequent calculus, is perhaps chiefly recognized as an early instance of a substructural logic, dating from 1958. As 
such, it precedes  Linear Logic~\cite{girardlinear} and the Bunched Implications~\cite{PymMono} logic underlying Separation Logic~\cite{reynoldslicssep}. See van Benthem's overview~\cite{vanbenthemaction} for a comparison to Linear Logic.

It  is interesting to note that the other main scourges of software security apart from command injection are memory corruption and unsafe resource usage, and that
substructural logics have been successful in reasoning about memory and resource usage~\cite{ishtiaqohearn,morrisettesop2000,reynoldslicssep,separationhiding}. 

Our view here of command injection as a kind of control effect that seizes its context evolved from calculi for continuations~\cite{danvyrepresenting, felleisenreasoning, felleisenprompt} and  
type-and-effect systems that make such control effects explicit in the types~\cite{griffin,polyeffects,controleffects,effectspopl03}. 
Behind each continuation, it is possible to introduce another level of continuations, sometimes called meta-continuations~\cite{danvyrepresenting}. These additional levels of continuations are particularly vivid in the syntactic calculus, as they are implicitly always present due to the silent double-negation introduction, without the need 
to write additional $\lambda$-abstractions.
  In linguistics, the double negation introduction is also known as ``type raising''. There are further examples of effects similar to those of control operators, such as Montague's semantics of quantification. For an introduction aimed at computer scientists, see Barker's survey article~\cite{barkercont}. 
  
For security policies or safety properties of programming languages, there are usually dynamic (run-time) and static (compile-time) approaches. 
A number of tools have been developed that defend against command injection attacks in a variety of languages~\cite{wassermanessence}. For such tools,
a major engineering challenge is to integrate them with existing technologies such as SQL and scripting languages with minimal intervention by programmers. While the use of parsing in such defences is one of the starting points of the present paper, the focus here is much more theoretical. 
Thiemann's Grammar-based Analysis of String Expressions~\cite{thiemanngrammar} uses a
language of types that appears closely related to the fragment of the Lambek calculus without implications $\lf$ and $\rf$. 

It remains a problem for future research to establish a formal connection between 
syntactic effects (such as those due to command injections) and control operators in the \emph{semantics} of the language, given by parsing actions~\cite{semparsing}. The semantic action of a string with a syntactic effect (such as those arising in command injections) may be conjectured to be equivalent to an expression with a suitable control operator, most likely a form of delimited continuation, such as shift/reset~\cite{danvy1990abstracting}.



\end{document}